\documentclass[twocolumn,prl,showpacs,superscriptaddress]{revtex4}
\usepackage{graphicx,amsmath}
\usepackage{color}

\begin{document}

\newcommand{\vect}{\vec}

\title{Photon correlation spectroscopy
  with incoherent light}

\author{D. Salerno}
\author{D. Brogioli}
\affiliation{Dipartimento di Medicina Sperimentale,
  Universit\`a degli Studi di Milano - Bicocca, Via Cadore 48, Monza
  (MI) 20052, Italy.}
\author{F. Croccolo}
\affiliation{Physics Department, University of Fribourg
  Perolles, 1700 Fribourg, Switzerland.}
\author{R. Ziano}
\author{F. Mantegazza}
\affiliation{Dipartimento di Medicina Sperimentale,
  Universit\`a degli Studi di Milano - Bicocca, Via Cadore 48, Monza
  (MI) 20052, Italy.}

\date{\today}

\begin{abstract}
Photon correlation spectroscopy (PCS) is based on measuring the temporal correlation of the light intensity scattered by the investigated sample. A typical setup requires a temporally coherent light source.
Here, we show that a short-coherence light source can be used as well,
provided that its coherence properties are suitably modified. This results in
a "skewed-coherence" light beam allowing that restores the coherence requirements. This approach overcomes the usual need for beam filtering, which would reduce the total brightness of the beam.
\end{abstract}

\maketitle


Scattering is a widespread phenomenon in which a radiation is forced to deviate from its original trajectory by the presence of inhomogeneities in the propagating medium \cite{omar,chaikin}. This phenomenon is common in several fields of science including particle physics, acoustics, geology, and plasma physics \cite{zemb}.
In optics, a particular application of this phenomenon, based on the evaluation of the temporal correlation functions among the scattered photons, is called PCS \cite{berne}. This technique has been an essential tool in soft matter physics especially for the study of colloids, liquid mixtures, gels, and polymers \cite{chu}.

PCS requires long temporal coherence of the impinging beam with respect to the light path differences of the scattered light to allow coherent homodyne superposition and therefore interference.
In this letter, we introduce a technique that uses short coherence radiation to perform PCS. It is based on the ``skewed-coherence'' beam \cite{martinez86,porras2003}, a particular arrangement of the optical field that exhibits a coherence neither spatial nor temporal, but along a well defined spatio-temporal trajectory.
The coherent regions of a skewed beam are thin disks (with radii equal to the transversal coherence length, and thicknesses equal to the longitudinal coherence length) that are propagating inclined (skewed) with respect to the normal of the Poynting vector of the beam.
This particular optical field has already been used in non-linear optics in combination with transform limit pulses, where the coherence region correspond to the intensity region, to avoid group velocity mismatch and walk off in $\chi^2$ crystals \cite{ditrapani1998}.
More recently it was shown that optical fields with skewed coherence (or its axial revolution, the X-waves) are spontaneously generated in parametric wave mixing processes \cite{picozzi2002,jedrkiewicz2006,jedrkiewicz2007}.
However, to our knowledge this is the first time that these kind of beams have been observed in connection with continuous emission and that possible applications to scattering measurements have been proposed.
In particular, we show that a short-coherence beam can be easily turned, through linear interaction, into a skewed coherence beam and that such coherence allows us to perform PCS experiments despite the short temporal coherence. We provide experimental evidence in the optical regime, and we propose the relevance of this technique for X-ray scattering.


\begin{figure}
\includegraphics[scale=0.6]{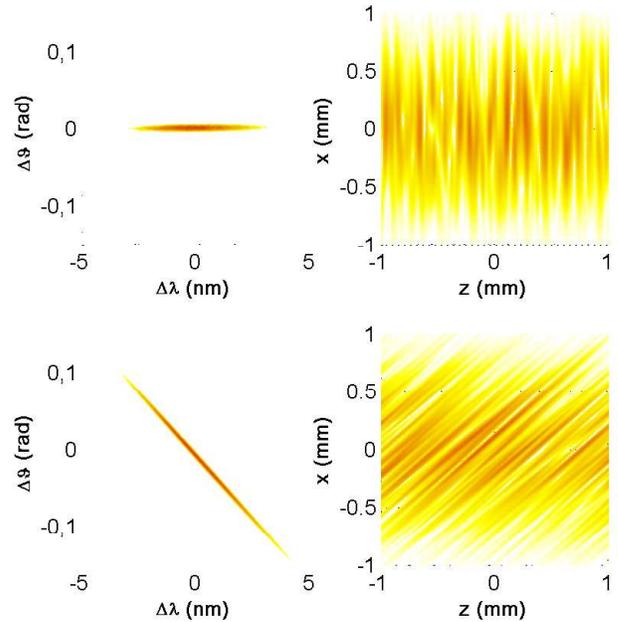}
\caption{
\label{fig:angdispersion}
Left column: measured $\Delta\theta-\Delta\lambda$ spectra of the non-skewed (first line) and skewed (second line) short coherence sources. Right column: realistic real space 2D intensity maps of the corresponding (same row) spectra calculated based on the data from the left column and the beam characteristics (see text for details). z is the propagating direction and x a generic transverse coordinate.
}
\end{figure}

We used a short-coherence laser diode source
(Sacher Lasertechnik, SAL-0660-025) emitting at 660nm, generally employed for external cavity applications \citep{conroy2000}. The laser diode was coated with an anti-reflective
covering on the front facet and driven under threshold.
To obtain the desired skewed coherence, the beam is diffracted by a reflection grating. When a short-coherence beam enters a reflection grating, it is dispersed into many diffraction orders. Considering the paths of the traveling waves, geometrical and continuity reasons impose that the propagating coherence regions of the diffracted beam are discs inclined by an angle $\sigma$ (skew angle) with respect to the normal of the direction of propagation \cite{borracz}. In our case, to obtain the desired skewed coherence, the beam is diffracted by a blazed reflective grating, with 600 lines per millimeter, blazed at $17.5^{\circ}$. The resulting skew angle $\sigma$ can be adjusted about from $20^{\circ}$ to $50^{\circ}$ by rotating the reflection grating and thereby changing the reciprocal angle between the incoming beam and the grating.

\begin{figure}
\includegraphics[scale=0.9]{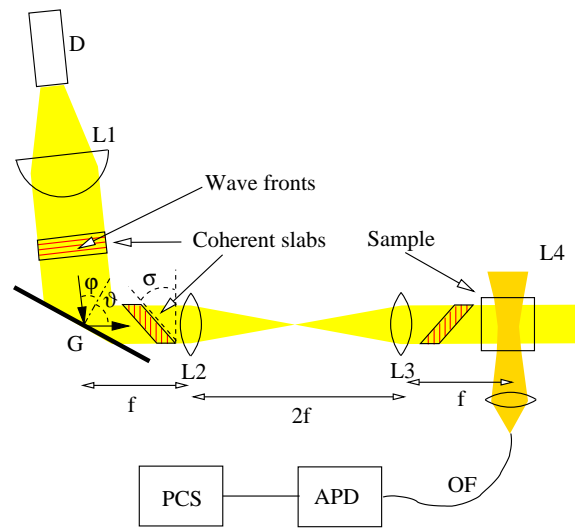}
\caption{
\label{fig:setup}
Experimental setup. D: diode;
L1: collimating aspheric lens;
G: blazed reflective grating, 600 lines per millimeter;
L2 and L3: telescope, with 1:1 magnification;
L4: collection optics;
OF: optical fiber;
APD: avalanche photodiode.
The short coherence beam impinges the grating G with the coherence regions oriented perpendicularly to the propagating direction ($\sigma$=0) and exits the grating with the skewed coherence regions oriented at an angle $\sigma$. See text for details.
}
\end{figure}

The diffracted and incoming beams follow the diffraction rules: for the first order, $\sin \vartheta - \sin \varphi = \frac{\lambda}{d}$ where $\vartheta$ and $\varphi$ are the angles between the normal to the grating of the diffracted and incoming beams, respectively, $\lambda$ is the wavelength, and $d$ is the grating spacing (see also Fig. \ref{fig:setup} for an illustration of the angles).
The skew angle $\sigma$ can be calculated through simple geometric considerations by imposing that the optical path traveled by the incoming and diffracted waves should be the same: $\tan \sigma = \tan \vartheta - \frac{\sin \varphi}{\cos \vartheta}$

We verified the coherence characteristics of the produced skewed beams by measuring the angular dispersion obtained at the output of reflection grating. To do that we acquired the images generated at the exit of a custom-made spectrograph in Czerny-Turner configuration. The two spatial coordinates of the images correspond to the $\vartheta - \lambda$ spectrum of the beam exiting from the grating.
Fig. \ref{fig:angdispersion} (left column) presents the $\Delta\vartheta-\Delta\lambda$ data measured from the recorded images in the case of a short coherence beam (obtained as zeroth order diffraction i.e, simple reflection) and a skewed short coherence beam (obtained as first order diffraction). The second column of Fig. \ref{fig:angdispersion} present a statistically realistic 2D section (transverse and propagation directions) of the intensity profile of the beam. The figure is obtained via FFT, by adding a random phase to the measured amplitude spectral fields and is scaled by using the data gathered from the angular dispersion measurements (temporal and spatial spectral width), the real dimension of the beam (continuous wave and diameter) and the temporal coherence measured by a interferometric experiment with a  Michelson-Morley set-up.

This representation clearly shows that the coherence regions of the beam correspond to the intensity spots where the different spectral components add up coherently.
In the case of zeroth order diffraction, where the beam maintains the same coherence properties of the impinging beam, we found, as expected, a significant large temporal spectrum (7~nm FWHM bandwidth) and a small spatial spectral bandwidth ($\sim$0.7 mrad) corresponding to a single transverse mode.

The coherence region of such a beam results as thin discs with radii of 0.9mm and thicknesses of $240 f$s (corresponding to about 75~$\mu$m, i.e. to an aspect ratio of the coherent regions of $\sim$ 12), and orientation perpendicular to the direction of propagation.
In the case of the first order of diffraction, the spectrum clearly shows the angular dispersion introduced by the reflection grating. This results in a coherence region oriented along a particular inclined direction in real space. The coherence regions are again slabs with the same aspect ratio of the zeroth order diffracted beam, but they are skewed by an angle $\sigma$ with respect to the perpendicular of the direction of propagation \citep{porras2003} .


Fig. \ref{fig:setup} shows a sketch of the PCS setup of the experiments we performed by illuminating the sample with the skewed beam. The short coherence beam passes through a collimating lens (L1) and is diffracted by the grating G.
A couple of lenses (L2 and L3) with focal $f=15~\mathrm{cm}$ are arranged in the $f-2f-f$ configuration (a telescope with magnification 1) conjugating a plane close to the grating to a plane inside the sample, which in turn is contained in a standard 1~cm square section glass cell.
The scattered light is collected using an optical system consisting of an optical fiber (OF) coupled with a GRIN lens (L4) placed at $90^{\circ}$ with respect to the impinging beam. The fiber output is then sent directly to the avalanche photodiode (APD) of a commercial PCS apparatus (ZetaPlus, Brookhaven Instruments), which is able to measure the time correlation function of the scattered intensity.

\begin{figure}
\begin{center}
\includegraphics[scale=0.6]{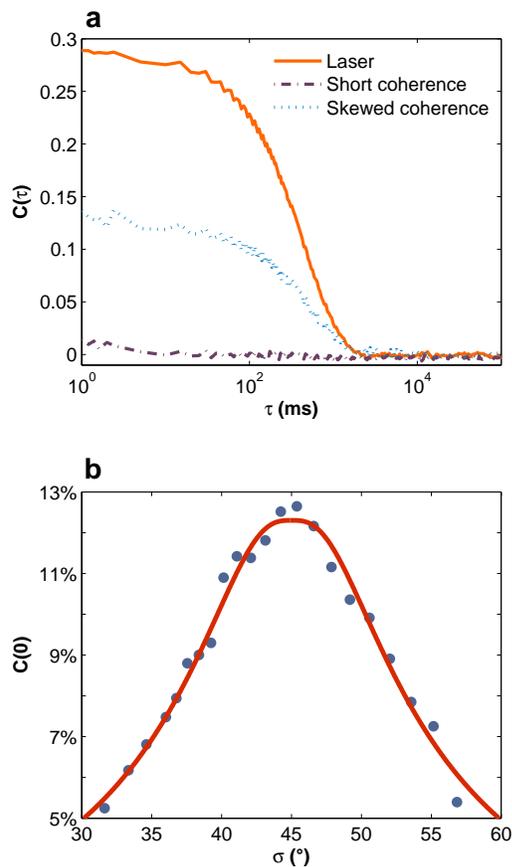}\\
\caption{
\label{fig:dls}
Panel (a): Measured PCS correlation function for a water suspension of 150~nm diameter colloids.
Data obtained with different coherence conditions of the illuminating beams.
Red line: laser beam; green dotted line: short-coherence beam; blue dotted
line: skewed coherence beam with $\sigma$ = $45^{\circ}$.
Panel (b): Contrast of the correlation function measured as a
function of the skew angle $\sigma$ for beams with different skewnesses and a detection angle of $90^{\circ}$.
The continuous line represents a fit with a theoretical
model, as explained in the text.}
\end{center}
\end{figure}

The first experiment aims to obtain the intensity temporal correlation function $C(\tau)=<I(t)I(t+\tau)>$ of the light scattered by a suspension of 150~nm diameter polystyrene particles (Duke scientific) in water at a volume fraction of $8X10^{-6}$.
Fig. \ref{fig:dls}a shows the normalized correlation functions obtained with different coherence conditions of the impinging beams.
When illuminating with laser light (Fig. \ref{fig:dls}a, continuous line), a classical exponential decay is obtained with a contrast close to 30\%. Such exponential decay is characteristic of the Brownian motion of mono dispersed nano-particles, and their size can be obtained using standard fitting procedures which provides a value very close to the manufacturer data (147nm).
With short-coherence light at $\sigma$=0 (Fig. \ref{fig:dls}a, dot-dashed line), the correlation function becomes completely flat. In reality a small contrast is still present in the order of 1\%, but the resulting correlation function is too noisy to be conveniently fitted and thus it is not possible to obtain any information about the sample size.
If the short-coherence beam is conveniently skewed, without applying any spectral filtering, a contrast of about 13\% is recovered (Fig. \ref{fig:dls}a, dotted line), enough to allow a precise evaluation of the decay time and thus of the particles sizes. The results of the fitting provide data similar to those obtained with the laser beam (145nm), thus showing that the correlation function is not dependent on the beam skewness other than in its contrast. In Fig. \ref{fig:dls}b we show the contrast of the measured autocorrelation function of the light intensity collected at a fixed scattering angle ($90^{\circ}$) as a function of the skew angle. The plot presented in Fig. \ref{fig:dls}b shows that the maximum contrast is actually achieved at a skew angle ($45^{\circ}$) equal to the half of the detection angle ($90^{\circ}$).

\begin{figure}
\begin{center}
\includegraphics[scale=0.5]{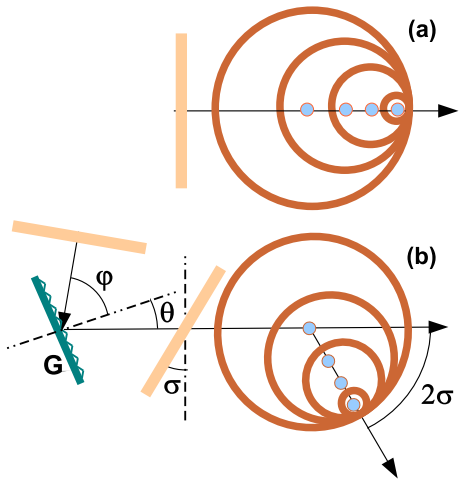}
\caption{
\label{fig:explanation}
Qualitative interpretation of the mechanism of scattering in the presence
of a short-coherence source (a) and a skewed source (b). The homodyne interference of the scattered light is possible only where the coherent regions overlap.
The thick stripe (yellow online) represents a coherent region of the primary beam, the dots (blue online) represent the diffusing particles that are able to produce interference, and the circular shells of increasing radius (red online) represent the coherence regions of the scattered spherical waves. See text for details.}
\end{center}
\end{figure}

In Fig. \ref{fig:explanation}, a two-dimensional schematic explanation of the obtained results is presented. The thick stripe (yellow online) corresponds to one of the coherent slabs of the impinging beam, the small dots (blue online) correspond to the scattering particles that contribute to interference, and the circles of increasing radius (red online) represent the scattered waves.
Such circles or scattered waves have a thickness equal to the temporal coherence of the impinging beam and have different radii because they are emitted at different times, each of them in phase with the impinging beam.
In the case of a non-skewed short-coherence beam, as generated by the undiffracted diode source, (Fig. \ref{fig:explanation}a) only for forward scattering is it possible to observe the homodyne interference between different diffusing waves. The scattered waves are generated when the impinging beam passes over the diffusing particles, so that all of the scattered waves are in phase with the impinging beam and consequently with each other along the propagation direction.
In Fig. \ref{fig:explanation}b the beam still has a short temporal coherence, but it is skewed by an angle $\sigma$, as obtained after reflection on the grating indicated by G in Fig. \ref{fig:explanation}. The coherent slab hits the various scattering objects at different times, thus the path length differences are compensated only if the scattering objects are located along an angle that is twice as large as the skewed angle.
It should be noted that the picture in Fig. \ref{fig:explanation}b is a slightly simplified 2D projection and that the real interference occurs on the surface of a cone; therefore, the exact angle at which the coherence condition is restored spans from $0^{\circ}$ to 2$\sigma$, depending on the azimuthal angle.
Practically speaking, the present technique takes advantage of the geometrical shape of the beam coherence to increase, over a direction specified by the skewed angle, the coherence volume (the volume of the sample that contributes to the interference signal) inside the sample. As shown in Fig. \ref{fig:explanation} b the coherence volume is oriented along an angle 2$\sigma$ with respect the propagation direction.

Thanks to this simplified picture, we are able to quantitatively model the dependence on the skewness angle $\sigma$ of the autocorrelation contrast (see Fig. \ref{fig:dls}b).
Indeed, the contrast of the correlation function depends on the coherent overlapping between the scattered waves. The correlation function for the short coherence beam ($\sigma$=0) is flat (contrast=C(0)=0) because the longitudinal coherence length is shorter than the optical path differences between the beams scattered from different regions of the sample. The reciprocal coherence of scattered waves is restored with a skewed beam at a detection angle equal to the twice the value of the beam skewness. The measured contrast shown in Fig. \ref{fig:dls}b can be theoretically derived assuming that the coherence volume is given by the geometrical intersection between the observed volume and a volume equal to the coherence slab but rotated to an angle $\sigma$.
The observed volume is the volume of the sample that is actually detectable, i.e., optically imaged on the photodetector, given the geometrical orientation of the collecting optics. Considering the skewed coherence slab as a disk or an oblate spheroid oriented with a variable angle $\sigma$, while the observed
volume is a fixed $45^{\circ}$ oriented stripe of thickness
comparable with the temporal coherence length, it is possible to extract the theoretical curve shown in Fig. \ref{fig:dls}b, by calculating the volume of the geometric intersection of the two regions.
By fitting the data with this model and using the actual aspect ratio of the skewed volume as the only free
parameter,
a value of the aspect ratio of about 11 is obtained, which is very close to the aspect ratio directly obtained via the spectrum and interferometric measurements shown above.

The phenomenon presented within this paper can be easily extended beyond the optical field to all the fields that share the same concept of spatial and temporal coherence. In particular its use for the X-ray regime can be of critical interest because the poor coherence characteristics (both spatial and temporal) of the available sources makes difficult to perform PCS experiments. In fact, PCS has been realized in the X-ray regime with synchrotron and FEL radiation (see \cite{mochrie1997}\cite{sutton2008} \citep{nugent2010} \citep{zanchetta2010} and references therein), that are the only sources with enough energy that can sustain the strong spatial and temporal filtering necessary to obtain the required transverse and temporal coherence.
Here, we suggest that a skewed X-ray beam can be obtained by impinging the X-ray beam on a Bragg crystal that naturally diffracts at different angle the different wavelength. This selects an angular dispersed beam that exhibits a skewed coherence in the real space. The resulting beam, suitable for X ray PCS, is significatively more bright of what can be obtained by applying subsequent spatial and temporal filtration.


In conclusion, we showed that modifying a short-temporal coherence beam into a skewed one, with coherence regions inclined with respect to the front wave, allows us to perform PCS with a short-coherence beam. A quantitative explanation of the observed phenomenon has been provided, and a good agreement with the experimental data has been reported.
Finally, we suggest that this method can be a real breakthrough in X-ray scattering where sources with the necessary coherence properties are not easily accessible and are very weak.

This work has been partially financially
supported by the EU (project NAD CP-IP 212043-2). FC acknowledges present support from EC Marie Curie
funding under grant IEF-251131, DyNeFl project.

\bibliography{coherence}

\end{document}